\documentclass[aps,prl,twocolumn,showpacs]{revtex4-1}
\usepackage{graphicx,amsmath,amssymb,bm}

\usepackage[usenames]{color}

\begin{document} 
 
\title{Static Response of Neutron Matter}

\author{Mateusz Buraczynski and Alexandros Gezerlis}
\affiliation{Department of Physics, University of Guelph, 
Guelph, ON N1G 2W1, Canada}

\begin{abstract} 
We generalize the problem of strongly interacting neutron matter by adding a periodic external modulation. 
This allows us to study from first principles a neutron system that is extended and inhomogeneous, 
with connections to the physics of both neutron-star crusts
and neutron-rich nuclei. We carry out fully non-perturbative microscopic Quantum Monte Carlo calculations of
the energy of neutron matter at different densities, as well as different strengths and periodicities of the external 
potential. In order to remove systematic errors, we examine finite-size effects and the impact of 
the wave function ansatz. We also make contact with 
energy-density functional theories of nuclei and
disentangle isovector gradient contributions from bulk properties. 
Finally, we calculate the static density-density linear response function of 
neutron matter and compare it with the response of other physical systems. 
\end{abstract} 

 
\maketitle 

Neutron matter (NM) is directly related to the properties of neutron-star crusts
and cores~\cite{Gandolfi:2015}.
The neutron-matter equation of state (EOS) has been intensively studied over the 
last few decades 
within \textit{ab initio} approaches~\cite{Friedman:1981,Akmal:1998,Schwenk:2005,Gezerlis:2008,Epelbaum:2008b,Kaiser:2012},
due to its astrophysical relevance and its connections with the physics of neutron-rich nuclei.
Neutron matter, though strongly interacting, is not self-bound, making it more straightforward
to tackle than isospin asymmetric matter or nuclei. This has 
led to the use of pure neutron systems as a testbed of nuclear forces, whether 
phenomenological~\cite{Carlson:Morales:2003,Gandolfi:2009,Gezerlis:2010,Gandolfi:2012,Baldo:2012} 
or chiral~\cite{Hebeler:2010,Tews:2013,Gezerlis:2013,Coraggio:2013,Hagen:2014,Gezerlis:2014,Carbone:2014,Roggero:2014,Wlazlowski:2014}.

While the interface between NM EOS calculations and neutron-star properties is relatively direct
(following from Einstein's field equations), the connection with 
neutron-rich nuclei is more circuitous.
It usually proceeds via nuclear energy-density
functionals (EDFs), the only approach currently able to provide a global description of the 
nuclear chart~\cite{Bender:2003}. 
EDFs contain a number of undetermined parameters, 
typically fit to nuclear masses and radii.
EDF approaches have benefitted from
including as constraints ``synthetic data'', i.e., dependable \textit{ab initio} many-body results.
Most commonly, 
these refer to the EOS of neutron matter~\cite{Fayans,SLy,Brown:2000,
Gogny,Fattoyev,Brown:2014,Rrapaj:2015},
though other quantities have also been used,
like the neutron pairing gap~\cite{Chamel:2008}, 
the energy of a neutron impurity~\cite{Forbes:2014,Roggero:2014b}, 
or the properties of artificially confined neutron drops~\cite{Pudliner:1996,
Pederiva:2004,Gandolfi:2011,Potter:2014}. 

In this Letter, we introduce another constraint: the static response of neutron matter, 
namely the behavior of an extended neutron system in the presence of a periodic external potential. Like neutron drops or 
the neutron polaron, the static-response problem provides input into the physics of 
neutron-rich nuclei. Furthermore, the static response of neutron matter
is directly equivalent to the situation prevailing
in a neutron-star crust. There, unbound neutrons interact strongly with each other \textit{and} with a lattice of nuclei: the unbound neutrons therefore experience the nuclei as a periodic modulation. 
The inhomogeneity of nucleon matter in nature has been addressed in several works,
employing a variety of approximations~\cite{Iwamoto:1982,Olsson:2004,Chamel:2011,Chamel:2013,
Kobyakov:2013,Davesne:2015}; this Letter opens up a corresponding \textit{ab initio} avenue.

The physical setting can be thought of as a periodic system of neutron drops.
Our investigation aims to provide insights on the relative magnitude of 
bulk vs surface effects. 
In other words, the static response of neutron matter can guide us
in the creation of ``neutron pasta''~\cite{Pethick:1995}, 
i.e., non-spherical nuclear systems, as well as their interplay with surrounding matter. 
Carrying out analogous calculations at very small densities one can then envision 
similarly disentangling pairing effects from the bulk and surface terms.

The problem of static response for strongly correlated
systems has a long history as well as direct experimental relevance, motivating us
to compare and contrast nuclear phenomena with systems in 
other areas of physics.
Specifically, there are close analogies with
both liquid $^4$He at vanishing pressure and temperature~\cite{Moroni:1992} and the three-dimensional 
electron gas~\cite{Moroni:1995}.
Similarly, small clusters of particles are experimentally accessible for 
cold Fermi gases in the presence of optical 
lattices~\cite{Pilati:2014,Carlson:2014}. Finally, there is an unmistakable connection to the prototypical problem of condensed-matter
physics: electrons interacting electromagnetically with each other and with a lattice of 
ions~\cite{Ashcroft}. 

Turning to the specifics of our investigation, we start from a microscopic Hamiltonian
containing a non-relativistic kinetic energy, a two-nucleon (NN) interaction  
(here taken to be the standard high-quality Argonne v8'
potential~\cite{Wiringa:2002}) and a three-nucleon (NNN) interaction 
(specifically, the Urbana IX potential~\cite{Pudliner:1997}). 
We will not focus on the details of the NN and NNN interactions in what follows.
The situation can be viewed as the application of 
a static (i.e., not dynamic) external potential $V_{\text{ext}}$
to an unperturbed Hamiltonian. Altogether, we have:
\begin{equation}
\hat{H} = -\frac{\hbar^2}{2m}\sum_i \nabla^{2}_{i} +\sum_{i<j}V_{ij}+\sum_{i<j<k}V_{ijk} + V_{\text{ext}}
\label{eq:ham}
\end{equation}
The last term is the external periodic potential: $V_{\text{ext}} = \sum_{i}V_i$, where $V_i=2v_q
\cos(\mathbf{q} \cdot \mathbf{r}_i)$. 
The periodicity of the potential is 
contained in $q$. We impose 
periodic boundary conditions on the wave function. 
If the density is $n$ and the 
simulation is carried out using $N$ particles, then the box length is $L = (N/n)^{1/3}$.
In order to respect translational invariance,
we ensure that an integer multiple of the potential's period fits in the box.

\begin{figure}[b]
\begin{center}
\includegraphics[width=\columnwidth]{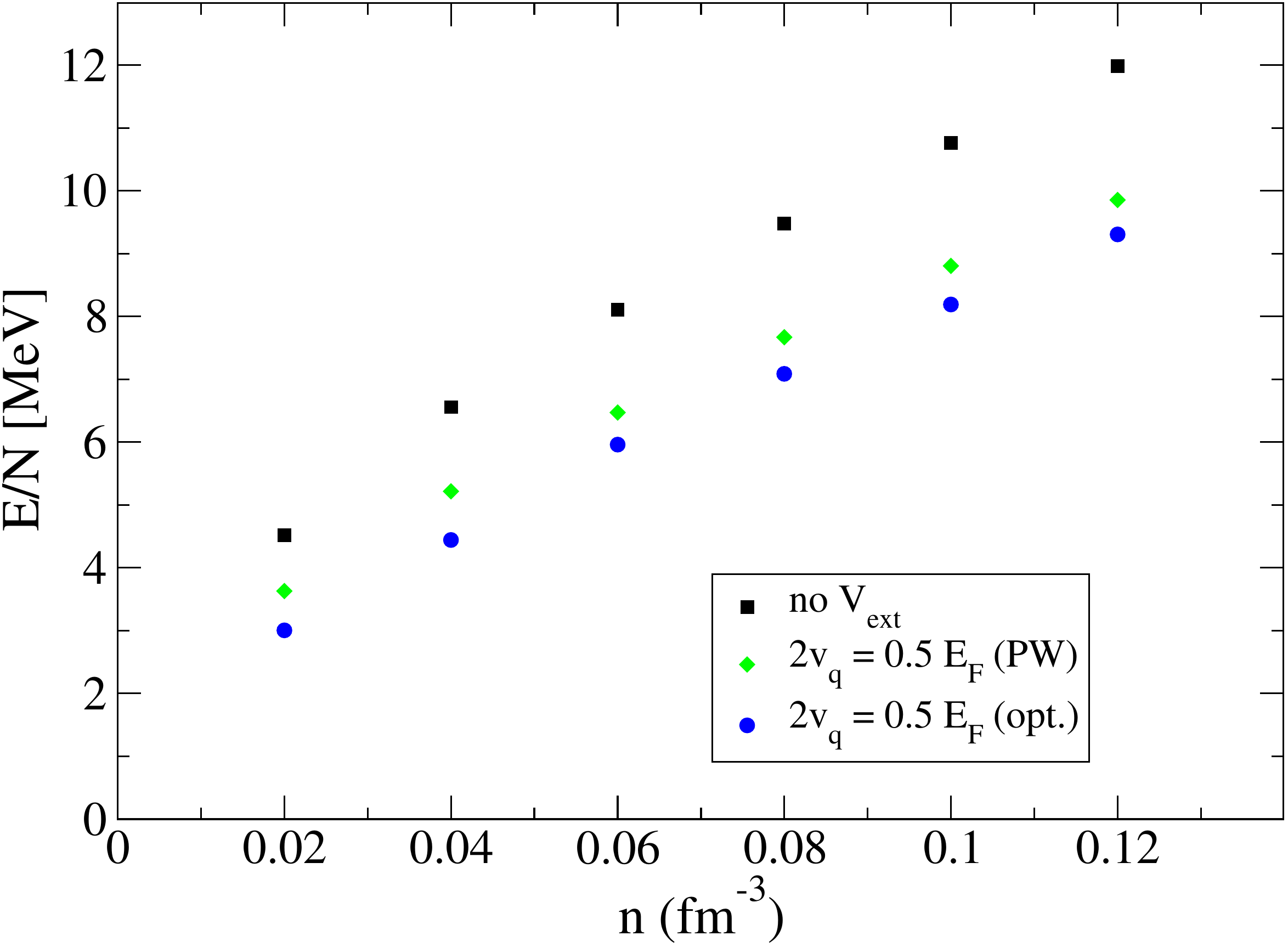}
\caption{(color online) Neutron-matter energy per particle as a function of density
using NN interactions and AFDMC.
Squares correspond to the case without a one-body potential; diamonds to a one-body
potential of fixed strength $2v_q = 0.5E_F$, periodicity $q=4\pi/L$,
and plane-wave single-particle orbitals (PW);
circles to a one-body potential of fixed strength $2v_q = 0.5E_F$, periodicity $q=4\pi/L$,
and optimized single-particle orbitals (opt.). 
\label{fig:ener_v_den}}
\end{center}
\end{figure}

We employ a family of \textit{ab initio} non-perturbative many-body 
approaches that have
been very successful in nuclear physics~\cite{Pudliner:1997,Gandolfi:2012,Gezerlis:2013},  
namely Quantum Monte Carlo (QMC). Specifically, we use Variational Monte Carlo (VMC),
which evaluates the expectation value of the Hamiltonian using a trial/variational 
wave function $| \Psi_T \rangle$, as well as Auxiliary-Field Diffusion Monte Carlo (AFDMC),
which extracts the ground state by projecting out excited states in imaginary
time $\tau$: $| \Psi_0 \rangle = \exp(-\hat{H}\tau)| \Psi_T \rangle$. Both methods 
have been applied to modern nuclear interactions, which include central, 
spin, tensor, spin-orbit, and more complicated terms. 

We used VMC and AFDMC to calculate the equation of state of neutron matter both 
excluding and including a modulating $V_{\text{ext}}$ potential. The results
for densities $0.02$ to $0.12$ ${\rm fm^{-3}}$ (roughly corresponding to the crust and the
outer core)
 are shown in 
Fig.~\ref{fig:ener_v_den}. Black squares show the result of using only an NN interaction
for 66 particles; they
match known values~\cite{Maris:2013}. 
(Here and below AFDMC results are shown 
-- the VMC results behave analogously~\cite{Buraczynski:2016}).
The trial wave function has the form:
\begin{equation}
| \Psi_T \rangle = \prod_{i<j} f(r_{ij})  ~{\cal A} \left [ \prod_{i} | \phi_i, s_i \rangle \right ]
\label{eq:PsiT}
\end{equation}
This is a product of a Jastrow factor (with central correlations, so in principle irrelevant)
and an antisymmetrized term (a Slater determinant of single-particle orbitals). 
As is standard for unpolarized neutron matter, these results follow from plane-wave orbitals
$\phi_i$.

Turning on the sinusoidal potential $V_{\text{ext}}$, one can expect that the energy
will be lowered: the particles will tend to stay away from repulsive regions and form 
clusters in the wells of the potential. We have carried out simulations that bear out
this expectation, using the same plane-wave orbitals as above, a periodicity $q=4\pi/L$, 
and a strength
that is a constant fraction of the Fermi energy $E_F$
($2v_q = 0.5E_F$, green diamonds in Fig.~\ref{fig:ener_v_den}). 
Note that our one-body strength is density dependent. Also, as the 
density increases, for fixed $N$ the box size $L$ decreases;
since the box length is a fixed multiple of the period, the latter decreases.
This is similar to what goes on in a neutron-star
crust, where as the density increases the lattice spacing decreases. 

We then proceeded to optimize the trial wave function $| \Psi_T \rangle $ in
VMC: instead of 
using plane-wave orbitals (which are the solution to the one-body problem in the absence
of an external potential), we employ Mathieu functions (the solutions to the one-body
problem in the presence of a sinusoidal potential). 
We also carried out a second optimization procedure: instead of assuming that the Mathieu
functions are eigenstates for noninteracting particles in the external potential with strength
$2v_q$, we used a new one-body strength $\beta$ 
(only in the evaluation of the wave function) as a variational parameter~\cite{Moroni:1992}. 
The final AFDMC result (using the $\beta$ that gives the minimum VMC energy) 
is shown using blue circles in Fig.~\ref{fig:ener_v_den}. The difference
between the plane-wave and Mathieu results (up to $\sim$1 MeV) underscores the importance
of optimizing the wave function properly. In comparison, the effect of the $\beta$
parameter is small, typically $\sim$50 keV.

\begin{figure}[t]
\begin{center}
\includegraphics[width=\columnwidth]{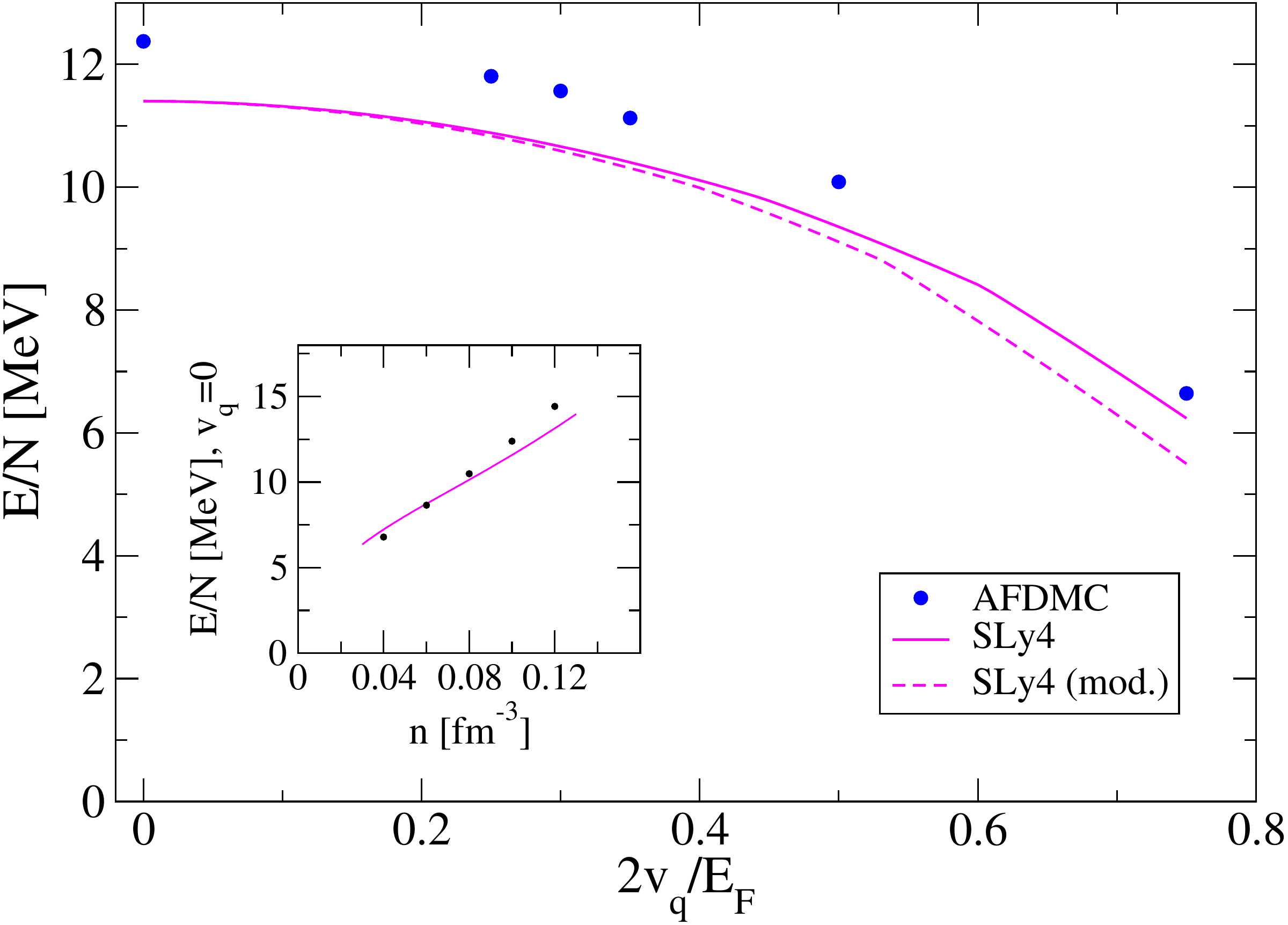}
\caption{Neutron-matter energy per particle as a function of one-body potential strength
at a density of $n=0.10\, {\rm fm^{-3}}$, using NN+NNN interactions and a 
one-body potential periodicity $q=4\pi/L$. Circles correspond to AFDMC results,
while the solid line follows from the SLy4 energy-density functional. 
The dashed line shows SLy4 results with a modified isovector gradient term.
Inset: AFDMC and SLy4 results in the absence of a one-body potential. 
\label{fig:ener_v_stren}}
\end{center}
\end{figure}

The periodically modulated EOS results in Fig.~\ref{fig:ener_v_den} 
were calculated using only
a single one-body potential strength, $2v_q = 0.5E_F$. While that was useful
in understanding the importance of the wave function ansatz, it 
may be misleading: there is, after all, nothing special about any one specific strength value.
Thus, it will be enlightening to study the dependence of the EOS on the 
potential strength, elucidating the departure from the case of homogeneous matter.
Figure~\ref{fig:ener_v_stren} shows the result of precisely such an investigation using
both NN and NNN interactions,
with $2v_q$ values chosen as follows: in order to ensure that the modulated ground-state
energy in VMC is statistically distinct from the homogeneous-matter energy,  $2v_q$  needs
to be not too small. At the other end, above a certain $2v_q$  the 
single-particle orbital filling starts to be considerably different from the homogeneous matter
(plane-wave orbital) case. We computed the energy for an open-shell case 
(70 particles) using different fillings of the single-particle orbitals, 
finding no appreciable change. As a result, we chose the 
one-body strengths $2v_q= \, 0.25, \, 0.30, \, 0.35, \, 0.50, \, 0.75$ times $E_F$.

In addition to AFDMC results denoted by circles, Fig.~\ref{fig:ener_v_stren} also shows 
results using the Skyrme SLy4 energy-density
functional~\cite{SLy} in the local-density approximation (solid line). 
As a reminder, 
the energy density for a large class of Skyrme EDFs can be given 
in the isospin $T$ representation:
\begin{equation}
{\cal E} = \sum_{T=0,1} \left [ ( C^{n,a}_T + C^{n,b}_T n^{\sigma}_0  ) n^2_T 
 + C^{\Delta n}_T n_T \Delta n_T + C^{\tau}_T n_T \tau_T  \right ]
\label{eq:Skyrme}
\end{equation}
along with an external potential-energy term (possibly also dependent on the isospin projection) 
$\sum_T n_T V_{\text{ext}, T}$, as well as a kinetic-energy term, which for simplicity can be taken to be an isoscalar: $\hbar^2\tau_0/2m$. 
The SLy4 $C$ values used to produce 
Fig.~\ref{fig:ener_v_stren} are one possible (standard) choice, though several others are being explored~\cite{Buraczynski:2016}.

At first sight, Fig.~\ref{fig:ener_v_stren} appears to be rather uninteresting: 
the AFDMC results are always more repulsive than the SLy4 ones, as found 
in QMC studies of neutron drops~\cite{Gandolfi:2011}. It is noteworthy, however, 
that the separation between the AFDMC and SLy4 results depends on the one-body
strength. Specifically, the isovector gradient 
term, $C^{\Delta n}_1 n_1 \Delta n_1$, which is absent in homogeneous neutron 
matter, has the 
effect of bringing the SLy4 results closer to the AFDMC ones as $2v_q$ 
is increased. 

Upon closer inspection, we realize that the more repulsive nature of the 
AFDMC results merely tracks the unperturbed ($v_q = 0$) 
relationship between AFDMC and SLy4 results (though 
at very large $2v_q$ the gradient term becomes strong enough to
change this). This crucial observation implies that, at different densities,
for many values of the one-body strength the modulated EOS under
SLy4 will have the same relationship to the AFDMC EOS 
as in the unmodulated ($v_q = 0$) case. This is illustrated in the inset
to Fig.~\ref{fig:ener_v_stren}, where the homogeneous neutron matter energy 
versus density 
in AFDMC and SLy4 is plotted. The AFDMC results 
are more repulsive at densities above 
0.06 fm$^{-3}$, while below that density the SLy4 ones are more repulsive. 
We have explicitly checked that the modulated interacting neutron system 
behaves as described at low density, 
for a variety of nuclear-force inputs.

Thus, the assumption that, since Skyrme functionals are fit to ab initio results for homogeneous matter, the difference between QMC and EDF in inhomogeneous neutron systems must come from the isovector gradient term alone~\cite{Gandolfi:2011}
is unwarranted.
In other words, at a finite one-body potential
strength the difference between QMC and EDF comes partly from the gradient term and 
partly from the bulk/homogeneous matter mismatch in energy. Thus, in order to separate
the two contributions while also respecting the homogeneous-matter properties, one should
modify the isovector gradient parameter $C^{\Delta n}_1$ 
so that the new Skyrme values match not the AFDMC ones, but 
a new set (here more attractive than the original) that is always as far away from 
the AFDMC perturbed value as the difference between the unperturbed AFDMC and SLy4 
energies. The result is the dashed line in Fig.~\ref{fig:ener_v_stren}; this 
required a modification to $C^{\Delta n}_1$ from -16 to -28 MeV fm$^5$,
using only the lowest-strength results as constraints in the new fit.

The density dependence shown in the inset to Fig.~\ref{fig:ener_v_stren},
taken together with our low-density findings mentioned above,
suggest a corresponding need for a 
density dependence in the adjustment to $C^{\Delta n}_1$. In other words,
if one is to carefully disentangle bulk from gradient contributions, the isovector 
gradient coefficient $C^{\Delta n}_1$ has to become density dependent 
(cf. the bulk term(s) 
$C^{n}$ in Eq.~(\ref{eq:Skyrme})), in addition to necessitating a global refit of all 
the Skyrme parameters. (This is consistent with
what is found using the density-matrix expansion~\cite{Holt:2011,Kaiser:2012b}.)
Another (possibly equivalent) option  is to consider gradient terms of different form~\cite{Carlson:2014}.

The calculations reported on in Figs.~\ref{fig:ener_v_den} \& \ref{fig:ener_v_stren} 
varied the density $n$
and one-body potential strength $2v_q$, respectively, but kept the one-body potential
periodicity $q$ fixed. We have also carried out calculations where the $q$ is varied
and thereby computed the linear density-density static response function of neutron 
matter. 

We first establish the notation~\cite{Pines:1966,Buraczynski:2016}. 
Let the unperturbed system be characterized by a Hamiltonian $\hat{H}_0$
(in our case, the first three terms in Eq.~(\ref{eq:ham})). 
The last term in Eq.~(\ref{eq:ham}) can then be viewed as the coupling 
of an external potential $V(\mathbf{r})$ to the one-body density operator $\hat{n} = \sum_i \delta(\mathbf{r}-\mathbf{r}_i)$: 
$\int d^3 r ~\hat{n}(\mathbf{r}) V(\mathbf{r})$, where $V(\mathbf{r}_i) \equiv V_i$.
We can now define the linear density-density static response function from the functional expansion of
the modulated density $n_{\text{tot}}(\mathbf{r})$ in terms of $V(\mathbf{r})$:
\begin{equation}
n_{\text{tot}}(\mathbf{r}) = n_0 + \int d^3 r' ~\chi(\mathbf{r}' - \mathbf{r}) V(\mathbf{r}')
\end{equation}
Explicitly, this $\chi(\mathbf{r}' - \mathbf{r})$ is the density-density response function of order 1, 
so a response function of order 3 
would involve three $V(\mathbf{r})$ terms, and so on. 
A similar expansion can be written down for the modulated 
energy (per particle) $E_{\text{tot}}/N$, which can then be re-expressed 
in terms of the Fourier components $v_q$ of the modulating potential, 
$V(\mathbf{r}) = \sum_{\mathbf{q}} v_{\mathbf{q}} \exp(i \mathbf{q} \cdot \mathbf{r})$, 
and the Fourier transform of the response function, $\chi(q)$:
\begin{equation}
\frac{E_{\text{tot}}}{N} = \frac{E_0}{N} + \frac{\chi(q)}{n_0}v^{2}_q+C_{4}v^{4}_q+ \cdots
\label{eq:ener_expan}
\end{equation}
The higher-order coefficients come from the higher-order response functions. 
Note that there are only
even powers present. It's also worth mentioning that the response function 
$\chi(q)$ is related to an energy integral over 
the dynamic structure factor.

\begin{figure}[b]
\begin{center}
\includegraphics[width=\columnwidth]{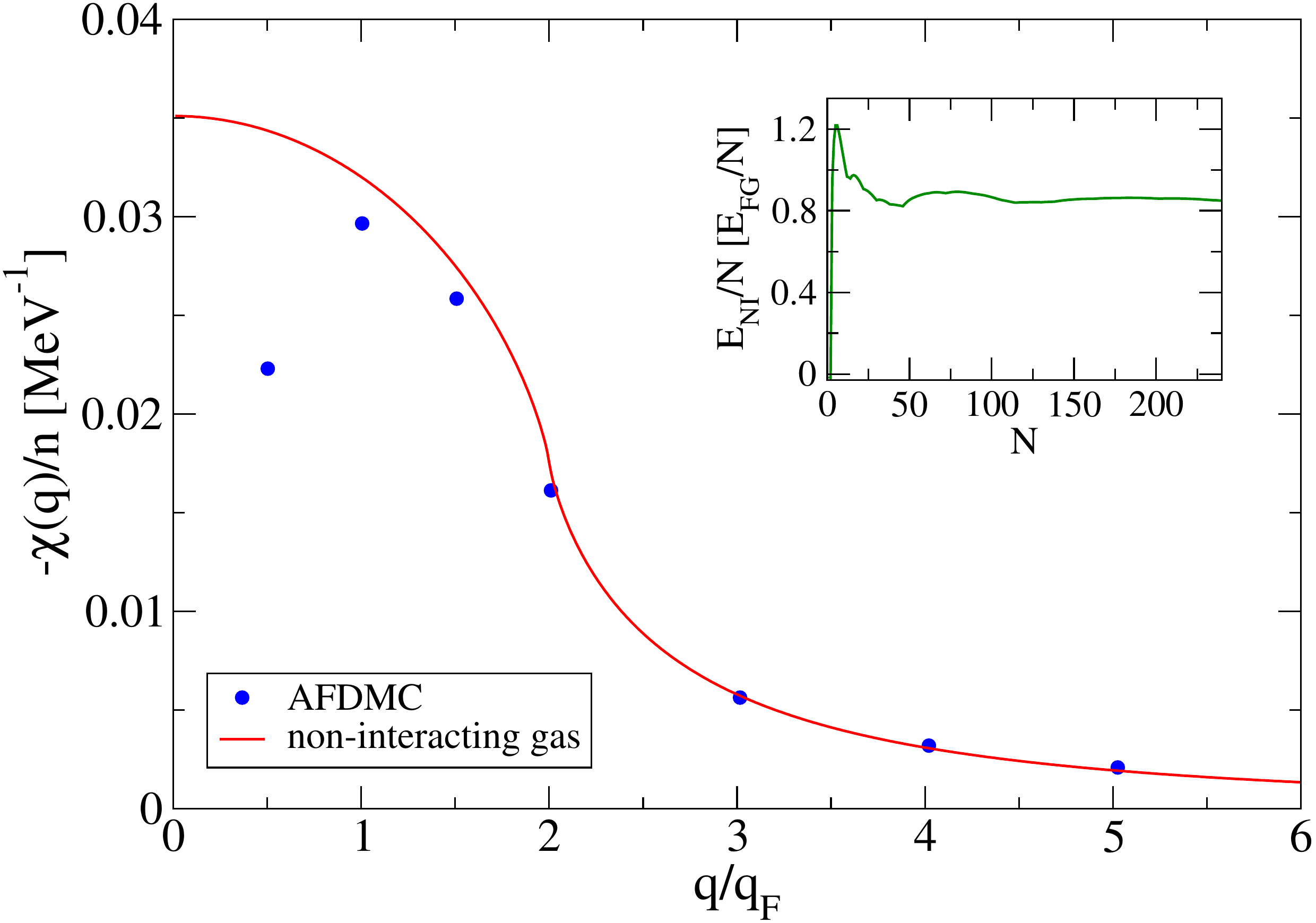}
\caption{Static-response function of 
neutron matter at a density of $n=0.10 \, {\rm fm^{-3}}$. 
Points follow from AFDMC results using 
NN+NNN interactions, as well as
several one-body strengths 
and periodicities. The line is 
the Lindhard function describing the response of a 
non-interacting Fermi gas. 
Inset: finite-size dependence of a non-interacting
Fermi gas in the presence of a one-body
potential of fixed strength $2v_q = 0.5E_F$ and periodicity $q=4\pi/L$.
\label{fig:response}}
\end{center}
\end{figure}

In producing the static-response function results 
we first addressed finite-size corrections.~\cite{Buraczynski:2016} The situation is illustrated in the inset 
to Fig.~\ref{fig:response}: removing NN and NNN interactions, one can compute 
quasi-analytically the energy of a modulated non-interacting Fermi gas,
$E_{NI}$. This is shown in units of 
the energy per particle in the unmodulated non-interacting Fermi gas, $E_{FG}/N = (3/5)E_F$.
The answer for 66 particles shows considerably larger shell effects than are present in the free Fermi gas. 
The thermodynamic-limit value depends on
the specific strength and periodicity of the potential being examined. We have 
carried out similar calculations with up to $\sim 100,000$ particles, arriving at a systematic way
of connecting finite-$N$ results to the thermodynamic limit. As a check, we 
adjusted the SLy4 results of Fig.~\ref{fig:ener_v_stren} and were able
to reproduce the thermodynamic-limit SLy4 values~\cite{SLy} to less than 0.5\%, 
thereby increasing our confidence in the prescription's validity.
(Since the periodic potential is externally specified, we do not expect major interaction-dependent finite-size effects, barring the extreme case of many clusters.).

For a fixed value of $q$ we carried out calculations 
for various one-body strengths 
($2v_q= \, 0.25, \, 0.30, \, 0.35, \, 0.50, \, 0.75$ times $E_F$,
as above)
and extracted the $\chi(q)$ using Eq.~(\ref{eq:ener_expan}) by 
fitting in powers of $v_q$ up to $v_q^4$. We repeated this process for 
several values of the periodicity: $q = 2, 4, 6, 8, 12, 16, 20$ times $\pi/L$.
This corresponds to 1, 2, 3, 4, 6, 8, 10 periods of the potential, respectively; note that even at 
the lowest density, this is still 2-3 times smaller 
than the relevant neutron-star crust scale.
The result of using NN+NNN interactions in AFDMC is shown in Fig.~\ref{fig:response}.
(Keeping different powers in Eq. (5) is related to how one
treats the higher-strength results, but the qualitative features are robust).
We reiterate that 
each point in this figure corresponds to an entire plot like Fig.~\ref{fig:ener_v_stren}. 
Thus, 
since each simulation uses a minimum of 66 strongly interacting particles, 
producing dependable \textit{ab initio} results for the static-response function 
of neutron matter requires 
a not insignificant amount of computing resources. Due to the varied connections with 
neutron-rich matter and nuclei, this effort is worthwhile.

Seeking a physical interpretation of our microscopic results, 
we compared them to the Lindhard function, corresponding to a non-interacting 
Fermi gas~\cite{Kittel:1969}:
\begin{equation}
\chi_{L} = - \frac{m  q_F}{2\pi^2 \hbar^2} \left [ 1 + \frac{q_F}{q} 
\left ( 1 - \left ( \frac{q}{2q_F} \right )^2 \right ) \ln \left | \frac{q + 2q_F}{q - 2q_F} \right |   \right ]
\label{eq:Lindhard}
\end{equation}
where $q_F$ is the Fermi wave number.
(We also used calculations like those in the inset to Fig.~\ref{fig:response} to reproduce this expression, 
further confirming our prescription.) 
We find that at large $q/q_F$ the interacting $\chi(q)$ 
matches the Lindhard function: since $C_4$ is small, Eq.~(\ref{eq:ener_expan}) shows
that when several clusters are present the interactions have a negligible effect. 
In the low-$q$ region our results exhibit a suppression: the particles are 
constrained closer to each other, thereby accenting the repulsion. 
This suppression is
different than in the electron gas, where 
at vanishing $q$ the response goes rapidly to zero~\cite{Moroni:1995}. 
We used the random-phase approximation~\cite{Pines:1966} 
to calculate the 
low-$q$ behavior for a model short-range interaction,
finding a value between the Lindhard and Coulomb results. While this result is merely qualitative,
it nicely parallels
our \textit{ab initio} prediction for the response of interacting neutron matter.

In summary, we have proposed the study of periodically modulated strongly interacting 
neutron matter. We have tackled this problem using Variational Monte Carlo and 
Auxiliary-Field Diffusion Monte Carlo. We investigated the ground-state energy
 at different densities,
paying close attention to the effect of optimizing the wave function. We then
addressed the question of how the homogeneous problem turns into the inhomogeneous one, 
by carrying out calculations for several values of the one-body potential strength. In that
process, we touched upon the effects our results have on Skyrme energy-density functionals
and disentangled the contributions of gradient and bulk terms. We then went on to study
several different one-body potential periodicities. After going over the basics of linear-response
theory, we were able to make a prediction for the static density-density response 
function of neutron matter. We found that the neutron-matter response function exhibits similarities to the 
response of the non-interacting Fermi gas and the electron gas, but also novel behavior. 
This work opens up a host of 
{\it ab initio} calculations for neutron-star crusts and improved descriptions of nuclei.

\begin{acknowledgments}
The authors are grateful to J. Carlson, S. Gandolfi, C. J. Pethick, and S. Reddy 
for many helpful discussions. 
This work was supported
by the Natural Sciences and Engineering Research Council (NSERC) of Canada and the 
Canada Foundation for Innovation (CFI). 
Computational resources were provided by SHARCNET and 
NERSC.
\end{acknowledgments}

\end{document}